\documentclass[a4paper]{article}
\usepackage{amscd,amsmath,amssymb}

\newcommand{\R}{\mathbb{R}}
\newcommand{\C}{\mathbb{C}}

\newcommand{\p}[1]{(\ref{#1})}

\newcommand{\cF}{{\cal F}}

\newcommand{\cD}{{\cal D}}

\newcommand{\cH}{{\cal H}}

\newcommand{\cW}{{\cal W}}

\newcommand{\cA}{{\cal A}}
\newcommand{\cB}{{\cal B}}
\newcommand{\cC}{{\cal C}}

\newcommand{\be}{\begin{equation}}
\newcommand{\ee}{\end{equation}}
\newcommand{\bea}{\begin{eqnarray}}
\newcommand{\eea}{\end{eqnarray}}

\newcommand{\ba}{\begin{array}} \newcommand{\ea}{\end{array}}

\newcommand{\sfrac}[2]{{\textstyle\frac{#1}{#2}}}

\newcommand{\nn}{\nonumber}

\begin{document}

\pagenumbering{gobble}

\begin{flushright}
ITP-UH-26/16
\end{flushright}\vspace{2cm}

\begin{center}
{\huge\bf Hidden symmetries of deformed oscillators}
\end{center}
\vspace{1cm}

\begin{center}
{\Large\bf  Sergey Krivonos${}^{a}$ , Olaf Lechtenfeld${}^{b}$
and Alexander Sorin${}^{a,c,d}$
}
\end{center}

\vspace{1cm}

\begin{center}
${}^a$ {\it
Bogoliubov  Laboratory of Theoretical Physics, JINR,
141980 Dubna, Russia}

\vspace{0.3cm}

${}^b$ {\it
Institut f\"ur Theoretische Physik and Riemann Center for Geometry and Physics \\
Leibniz Universit\"at Hannover,
Appelstrasse 2, D-30167 Hannover, Germany}

\vspace{0.3cm}
${}^c$ {\it
National Research Nuclear University MEPhI (Moscow Engineering Physics Institute) \\
Kashirskoe Shosse 31, 115409 Moscow, Russia}

\vspace{0.3cm}

${}^d$ {\it
Dubna International University, 141980, Dubna,
Russia}

\vspace{0.8cm}

{\tt krivonos@theor.jinr.ru, lechtenf@itp.uni-hannover.de, sorin@theor.jinr.ru}
\end{center}
\vspace{2cm}

\begin{abstract}\noindent
We associate with each simple Lie algebra a system of second-order differential equations 
invariant under a non-compact real form of the corresponding Lie group.
In the limit of a contraction to a Schr\"odinger algebra, these equations reduce to a system 
of ordinary harmonic oscillators. We provide two clarifying examples of such deformed oscillators: 
one system invariant under SO$(2,3)$ transformations, and another system featuring $G_{2(2)}$ symmetry. 
The construction of invariant actions requires adding semi-dynamical degrees of freedom;
we illustrate the algorithm with the two examples mentioned.
\end{abstract}

\newpage
\pagenumbering{arabic}
\setcounter{page}{1}
\section{Introduction}
It is widely believed that integrability of a mechanical system is related with a high degree of (usually hidden) symmetry.
Identifying such symmetry for a given system may be very complicated, even in the simplest cases, like in harmonic oscillators. 
The inverse task -- constructing a system possessing a given symmetry -- seems to be more simple,
since there are many ways to find its equations of motion. One of them is the method of nonlinear realizations~\cite{coset1, coset2}, 
equipped with the inverse Higgs phenomenon \cite{ih}. For constructing a system of equations with a given symmetry,
all one needs is the symmetry group together with the stability subgroup, which acts linearly on the mechanical coordinates.

Our recent paper \cite{KLS} applies nonlinear realizations to the Schr\"{o}dinger and $\ell$-conformal Galilei algebra.
These symmetries give rise to a system of ordinary harmonic oscillators and their higher-derivative (in time) extensions known as conformal
Pais--Uhlenbeck oscillators \cite{confPU,confPU1}. Next to the one-dimensional conformal algebra $so(1,2)\sim su(1,1)$, there are
only evident shift symmetries of the oscillators. However, when we deform the Schr\"{o}dinger algebra in two space dimensions
to $su(1,2)$, the corresponding oscillator is also deformed to a nonlinear one, but remains dynamically equivalent to the standard oscillator~\cite{KN}.
This suggests the existence of $F$-invariant nonlinearly deformed oscillator systems for every noncompact real Lie group~$F$.

Crucial in our construction~\cite{KLS} of the deformed oscillators is the 5-grading of $su(1,2)$. Now, {\it any\/} finite-dimensional 
simple complex Lie algebra beyond $sl_2$ has at least one non-compact real form with a 5-graded decomposition~\cite{5gr,grading}.
A universal part of the 5-grading is the $su(1,1)$ sub-algebra formed by the highest- and lowest-grade subspace together with 
the (grade-zero) grading operator~$L_0$, so one-dimensional conformal symmetry is always present.
In the present paper, we extend the procedure developed in~\cite{KLS} from $su(1,2)$ to a non-compact real form of any simple Lie algebra.
It will provide a system of (generically nonlinear) second-order differential equations with the prescribed non-compact symmetry,
which reduces to ordinary harmonic oscillators under the contraction to a Schr\"odinger algebra.

The existence of a corresponding invariant action is a more delicate matter, which we also investigate here.
It is not guaranteed, because the equations of motion usually enjoy a larger symmetry than the action.
In the following, we shall work out two explicit examples in detail, featuring $SO(2,3)$ and $G_{2(2)}$ symmetry, respectively.
We shall see that the formulation of an action requires additional, semi-dynamical degrees of freedom which, however, 
do not affect the deformed oscillator equations. This provides an algorithm for the construction of invariant actions.

\setcounter{equation}{0}
\section{General construction}
It is a well known fact \cite{5gr,grading} that every simple Lie algebra $\cF$ (except for $sl_2$) admits 5-graded decompositions with respect
to a suitable generator $L_0 \in \cF$:
\be\label{5gr}
\cF = \mathfrak{f}_{-1} \oplus  \mathfrak{f}_{-\frac{1}{2}} \oplus \mathfrak{f}_{0} \oplus  \mathfrak{f}_{+\frac{1}{2}}\oplus  \mathfrak{f}_{+1}
\qquad\textrm{with}\qquad
\left[  \mathfrak{f}_i,  \mathfrak{f}_j\right] \subseteq  \mathfrak{f}_{i+j} 
\quad\textrm{for} \ i,j\in\left\{ -1,-\sfrac{1}{2},0,\sfrac{1}{2},1\right\}
\ee
($\mathfrak{f}_i=0$ for $|i|>1$ understood).
There is an (up to automorphisms) unique 5-grading with one-dimensional spaces $\mathfrak{f}_{\pm1}$. Choosing this one, we may write
\be
\mathfrak{f}_{-1}=\C\,L_{-1}, \qquad \mathfrak{f}_{+1}=\C\,L_1 \qquad\textrm{and}\qquad \mathfrak{f}_0 = \cH \oplus \C\,L_0,
\ee
where $\cH \subset \cF$ is a Lie subalgebra and $L_0$ commutes with $\cH$. 
A basis for the spaces $\mathfrak{f}_{\pm\frac12}$ (of some dimension~$d$) is given by generators $G_{\pm\frac12}^A$ with $A=1,\ldots,d$.
They carry an irreducible representation of~$\cH$.
In the following, we will deal with {\it real\/} Lie algebras and groups only, so some real form of $\cF$ and $\cH$ has to be picked.
(We keep the same notation however.) Compatibility with the 5-grading requires this real form to be non-compact.
Therefore, $(L_{-1},L_1,L_0)$ generate an $su(1,1)$ subalgebra of $\cF$. 
Different real forms of $\cF$ and $\cH$ give rise to different
non-compact quaternionic symmetric spaces~$W$~\cite{5gr,grading},
\be\label{W}
W = \begin{array}{c} F \\ \hline
H\times \textrm{SU}(1,1)
\end{array},
\ee
where $F$, $H$ and SU(1,1) are the (simply-connected) groups generated by $\cF$, $\cH$ and $su(1,1)$, respectively.

The main idea of our construction consists in enlarging the coset by slightly reducing the stability group from 
$H\times \textrm{SU}(1,1)$ to $H\times\mathfrak{B}_{\textrm{SU}(1,1)}$, where $\mathfrak{B}_{\textrm{SU}(1,1)}$ denotes the positive 
Borel subgroup of SU(1,1), whose algebra $\mathfrak{b}_{su(1,1)}$ is generated by $(L_0, L_1)$. In other words, 
we keep $L_{-1}$ in the numerator and consider the coset
\be\label{cW}
\cW = \begin{array}{c} F \\ \hline
H\times \mathfrak{B}_{\textrm{SU}(1,1)}
\end{array}.
\ee
The elements of $\cW$ can be parametrized as follows,
\be\label{g1}
g=e^{t \left( L_{-1}+ \omega^2 L_1\right)} e^{u(t) \cdot G_{-\frac{1}{2}}} e^{v(t) \cdot G_{\frac{1}{2}}} ,
\ee
where we employed a $\cdot$ notation to suppress the summation over~$A$.
The parameter $\omega$ represents some freedom in the parametrization of $\cW$.
It yields the oscillation frequency of the deformed oscillators we are going to construct.

Defining the Cartan forms in the standard way (with a basis $\{h_s\}$ of $\cH$),
\be\label{cf1}
g^{-1} d g = \omega_{-1} L_{-1} +  \omega_{0} L_0 + \omega_{1} L_1 + 
\omega_{-\frac{1}{2}} \cdot G_{-\frac{1}{2}} + \omega_{\frac{1}{2}} \cdot G_{\frac{1}{2}} +
\sum_s \omega_h^s\,h_s,
\ee
one can check that the constraints
\be\label{ih1}
 \omega_{-\frac{1}{2}}=0
\ee
firstly are invariant under the whole group $F$, realized by left multiplication in the coset $\cW$ \p{cW}, and secondly 
express the Goldstone fields $v(t)$ through the Goldstone fields $u(t)$ and their time derivatives in a covariant fashion 
(inverse Higgs phenomenon \cite{ih}).
After imposing the constraints \p{ih1} we have a realization of the $F$ transformations on the time $t$ and the $d$ coordinates $u_A(t)$.

Finally, one can impose the additional invariant constraints
\be\label{eom1a}
\omega_{\frac{1}{2}}=0,
\ee
which produces a system of second-order differential equations for the variables $u_A(t)$. These are the equations of motion.
Hence, with every simple Lie algebra $\cF$ one may associate a system of dynamical equations in $d$ variables 
which is invariant under some non-compact real form of the group $F$.

Given the above structures, we can partially fix the commutator relations of~$\cF$:
\be\label{alg1}
\left[ L_n, L_m \right] = (n{-}m) L_{n+m}, \qquad 
\left[ L_n, G^A_r\right] = \left( \sfrac{n}{2}{-}r\right) G^A_{n+r}, \qquad 
m,n = -1,0,1, \  r= -\sfrac{1}{2},\sfrac{1}{2}, \  A =1,\ldots,d.
\ee
The $[G,G]$ commutators lands in $\cH\oplus su(1,1)$. However, they can be made to vanish by a group contraction.
To this end, one rescales the generators via $G{}^A_{\pm \frac{1}{2}} = \gamma^{-1}{\widetilde G}{}^A_{\pm \frac{1}{2}}$ with $\gamma\in\R_+$.
The limit $\gamma \rightarrow 0$  preserves the relations \p{alg1} but lets all generators ${\widetilde G}{}^A_{\pm \frac{1}{2}}$ commute with one another.
Thus, after the contraction we arrive at the algebra
\bea\label{confGal}
\left[ L_n, L_m \right] = (n{-}m) L_{n+m}, \qquad 
\left[ L_n, {\widetilde G}{}^A_r\right] = \left( \sfrac{n}{2}{-}r\right) {\widetilde G}{}^A_{n+r}, \qquad
\left[ {\widetilde G}{}^A_{\pm \frac{1}{2}}, {\widetilde G}{}^B_{\pm \frac{1}{2}}\right] =0, \qquad
\left[ {\widetilde G}{}^A_{\pm \frac{1}{2}}, {\widetilde G}{}^B_{\mp \frac{1}{2}}\right] =0.
\eea
This is the Schr\"{o}dinger algebra in $d{+}1$ dimensions~\cite{KLS}. 
One may check that in this limit the equations \p{ih1} and \p{eom1a} linearize to
\be\label{ho}
\ddot{u}{}_A(t) + \omega^2  u{}_A(t) =0 \qquad\textrm{for}\quad  A =1, \ldots, d.
\ee
Undoing the contraction, one may regard \p{ih1} and \p{eom1a} as a deformation of \p{ho}.
For this reason we refer to them as `deformed oscillators'.
The first example, for the algebra $\cF = su(1,2)$ and $\cH=u(1)$, was considered in \cite{KLS} and~\cite{KN}.

Finally we note that the above construction yields only the equations of motion for the variables $u_A(t)$. 
The question of existence of a corresponding invariant action has to be answered independently. 
We will demonstrate below that a positive answer requires extending further the number of Goldstone fields.

In the following two sections we will consider two instructive examples in detail: $SO(2,3)$ and $G_{2(2)}$ invariant deformed oscillators.

\setcounter{equation}{0}
\section{$SO(2,3)$ invariant oscillator}
The 10-dimensional $so(2,3)$ algebra admits a 5-graded structure with $d=2$ and $\cH=su(1,1)$. 
It can be visualized by writing the commutator relations as
\bea\label{so23alg}
&& \left[ L_n, L_m \right] = (n{-}m) L_{n+m}, \quad \left[ M_a, M_b \right] = (a{-}b) M_{a+b},\quad m,n =-1,0,1,\quad a,b =-1,0,1, \nn \\
&& \left[ L_n, G_{r,A}\right] = \left( \sfrac{n}{2}{-}r \right)  G_{n+r,A}, \quad
\left[ M_a, G_{r,A}\right] = \left( \sfrac{a}{2}{-}A \right)  G_{r,a+A},\quad r,s =-\sfrac{1}{2},\sfrac{1}{2}, \quad A, B =-\sfrac{1}{2}, \sfrac{1}{2},\quad  \nn \\
&& \left[ G_{r,A}, G_{s,B} \right] = 2\, \left( A\,  \delta_{A+B,0} \, L_{r+s} + r \, \delta_{r+s,0} \, M_{A+B}\right).
\eea
All generators may be taken to be antihermitian,
\be\label{cr1}
\left( L_n\right)^\dagger = - L_n,\qquad \left( M_a\right)^\dagger =  - M_{a},\qquad \left( G_{r,A}\right)^\dagger = - G_{r,A}.
\ee
Thus, we see that
\be
 \mathfrak{f}_{-\frac{1}{2}} = \R\,G_{-\frac12,-\frac12} \oplus \R\,G_{-\frac12,+\frac12},\quad  
 \mathfrak{f}_{+\frac{1}{2}}= \R\,G_{+\frac12,-\frac12} \oplus \R\,G_{+\frac12,+\frac12},\quad
 \cH = \R\,M_{-1}+\R\,M_0+\R\,M_{+1}.
\ee
{}From the maximally non-compact four-dimensional quaternionic symmetric space
$W=\textrm{SO}(2,3)/\textrm{SO}(2,2)$ we pass to the five-dimensional coset space
\be
\cW = \frac{\textrm{SO}(2,3)}{\textrm{SU}(1,1)\times\mathfrak{B}_{\textrm{SU}(1,1)}}
\ee
where the stability subgroup is generated by $( L_0, L_1, M_a)$.
The coset $\cW$ is parametrized
\be\label{g2}
g= e^{t (L_{-1}+\omega^2 L_1)} e^{ u_1 G_{-\frac{1}{2},-\frac{1}{2}}+u_2  G_{-\frac{1}{2},+\frac{1}{2}}}
e^{ v_1 G_{+\frac{1}{2},-\frac{1}{2}}+v_2  G_{+\frac{1}{2},+\frac{1}{2}}}, \qquad u_{1,2}^\star=u_{1,2}, \; v_{1,2}^\star=v_{1,2},
\quad g^\dagger = g^{-1}.
\ee
To find the equations of motion for the coordinates $u_1(t)$ and $u_2(t)$ we have to calculate the Cartan forms
\be\label{cf2}
g^{-1}dg = \sum_{n}\omega_{{L}_n}\, L_n+\sum_{a} \omega_{{M}_a}\, M_a+ \sum_{r,A} \omega_{r,A} G_{r,A}.
\ee
Their explicit form reads (we will not need $\omega_{L_n}$)
\bea\label{Cf2}
\omega_{-\frac{1}{2},-\frac{1}{2}} &=& d u_1 -v_1 \, \left( dt +\sfrac{1}{2} \left( u_1\, d u_2 -u_2\, d u_1\right)\right), \nn \\
\omega_{-\frac{1}{2},+\frac{1}{2}} &=& d u_2 -v_2 \, \left( dt +\sfrac{1}{2} \left( u_1\, d u_2 -u_2\, d u_1\right)\right),  \nn   \\
\omega_{+\frac{1}{2},-\frac{1}{2}} &=& d v_1 +\sfrac{1}{2} v_1 \left( v_2\, d u_1  - v_1\, d u_2\right)\,+\omega^2\, dt\, u_1
\left( 1 +\sfrac{1}{2}\left (u_2\, v_1 -u_1\, v_2\right)\right),\nn\\
\omega_{+\frac{1}{2},+\frac{1}{2}} &=& d v_2 +\sfrac{1}{2} v_2 \left( v_2\, d u_1  - v_1\, d u_2\right)\,+\omega^2\, dt\, u_2
\left( 1 +\sfrac{1}{2}\left (u_2\, v_1 -u_1\, v_2\right)\right)
\eea
and
\bea\label{mforms2}
\omega_{{M}_{-1}} & = &+\sfrac{1}{4}\, u_1\, v_1^2\, du_2 -v_1 \left( 1 +\sfrac{1}{4} u_2 v_1\right) du_1 +\sfrac{1}{2} dt \left( v_1^2 + \omega^2\, u_1^2\right), \nn \\
\omega_{{M}_{+1}} & = &-\sfrac{1}{4}\, u_2\, v_2^2\, du_1 -v_2 \left( 1 -\sfrac{1}{4} u_1 v_2\right) du_2 +\sfrac{1}{2} dt \left( v_2^2 + \omega^2\, u_2^2\right), \nn  \nn \\
\omega_{{M}_0} & = &  -v_2 \left( 1 +\sfrac{1}{2} u_2\, v_1\right) du_1-v_1 \left( 1 -\sfrac{1}{2} u_1\, v_2\right) du_2+
 dt \left( v_1\, v_2  + \omega^2\, u_1\, u_2\right).
\eea

In accordance  with the general scheme outlined in Section 2, we firstly have to express $v_1$ 
and $v_2$ in terms of $u_1$ and $u_2$
by nullifying the forms $\omega_{-\frac{1}{2},-\frac{1}{2}}$ and $\omega_{-\frac{1}{2},+\frac{1}{2}}$. 
Doing so, we  obtain
\be\label{ih2}
\omega_{-\frac{1}{2},\pm\frac{1}{2}}=0 \qquad \Rightarrow \qquad
v_{1}=\frac{\dot{u}_{1}}{1+\frac{1}{2}\left( u_1 \dot{u}_2 - \dot{u}_1 u_2\right)}
\qquad\textrm{and}\qquad
v_{2}=\frac{\dot{u}_{2}}{1+\frac{1}{2}\left( u_1 \dot{u}_2 - \dot{u}_1 u_2\right)}.
\ee
Finally, to find the invariant equations of motion one has to nullify the forms 
$\omega_{+\frac{1}{2},-\frac{1}{2}}$ and $\omega_{+\frac{1}{2},+\frac{1}{2}}$
(with \p{ih2} taken into account). In this way we arrive at
\be\label{eom2}
\omega_{+\frac{1}{2},\pm\frac{1}{2}}=0\qquad \Rightarrow \qquad
\ddot{u}_1+\omega^2\, u_1 =0 \qquad\textrm{and}\qquad \ddot{u}_2+ \omega^2\, u_2=0.
\ee
Having expected two coupled nonlinear differential equations, 
we are surprised to have obtained just ordinary (decoupled) linear oscillator equations,
even before taking the linearizing contraction limit to the Schr\"odingier algebra.
We conclude that the equations of motion of the ordinary two-dimensional harmonic oscillator 
enjoy an SO(2,3) invariance! 

It is instructive to present the SO(2,3) symmetry transformations of~\p{eom2} explicitly.
The SO(2,3) group is realized by left multiplication on the coset element \p{g2},
\be\label{left}
g_0\, g = g'\, h \qquad\textrm{with}\qquad
g_0 \in \textrm{SO}(2,3) \qquad\textrm{and}\qquad
h \in \textrm{SU}(1,1) \times \mathfrak{B}_{\textrm{SU}(1,1)}.
\ee
Different elements $g_0$ effect different changes $g\mapsto g'$, which induce transformations
of the time~$t$ and the coordinates $(u_1,u_2,v_1,v_2)$. We display their infinitesimal versions
(linear in the transformation parameters):\footnote{
The result of acting with 
$g_0=e^{\varepsilon_1\, G_{+\frac{1}{2},-\frac{1}{2}} + \varepsilon_2 \,G_{+\frac{1}{2},+\frac{1}{2}}}$ 
can be obtained from the commutator of \p{conf2} and \p{Gtr2}.}
\goodbreak
\bea
g_0=e^{a\, L_{-1}+b\, L_0 +c\,L_{+1}}:
&& \left\{ \begin{array}{lll}
\delta t&=&\frac{1+\cos(2 \omega t)}{2} \, a +\frac{\sin( 2 \omega t)}{2 \omega}\, b +\frac{1-\cos(2 \omega t)}{2 \omega^2}\, c\ \equiv\ f(t)\\
\delta u_1 &=& \frac{1}{2} \dot{f} \, u_1\\
\delta u_2 &=& \frac{1}{2} \dot{f} \, u_2 \\
\delta v_1 &=& -\frac{1}{2} \dot{f}\,v_1 +\frac{1}{2} \ddot{f} \, u_1\, \left( 1 + \frac{1}{2} \left( u_2\,v_1 -u_1 \, v_2\right)\right)\\
\delta v_2 &=& -\frac{1}{2} \dot{f}\,v_2 +\frac{1}{2} \ddot{f} \, u_2\, \left( 1 + \frac{1}{2} \left( u_2\,v_1 -u_1 \, v_2\right)\right)
\end{array}\right\} , \label{conf2} \\
g_0=e^{ \alpha \, M_{-1} + \beta \, M_0 + \gamma\, M_{+1}}: 
&& \left\{ \begin{array}{lll}
\delta t &=& 0 \\
\delta u_1 &=& \frac{1}{2}\, \beta\, u_1 - \alpha\, u_2 \\
\delta u_2 &=& -\frac{1}{2}\, \beta\, u_2 + \gamma\, u_1\\
\delta v_1 &=& \frac{1}{2}\, \beta\, v_1 - \alpha\, v_2 \\ 
\delta v_2 &=& -\frac{1}{2}\, \beta\, v_2 + \gamma\, v_1
\end{array} \right\} ,\label{m2} \\
g_0=e^{\epsilon_1\, G_{-\frac{1}{2},-\frac{1}{2}} + \epsilon_2 \,G_{-\frac{1}{2},+\frac{1}{2}}}:
&& \left\{ \begin{array}{lll}
\delta t &=& \frac{1}{2} \cos(\omega t) \left( \epsilon_2\, u_1 -\epsilon_1\, u_2\right) \\
\delta u_1 &=&  \cos(\omega t) \epsilon_1 -\frac{1}{2} \omega \sin(\omega t) \, u_1\, \left( \epsilon_2 u_1 -\epsilon_1 u_2\right)\\
\delta u_2 &=& \cos(\omega t) \epsilon_2 -\frac{1}{2} \omega \sin(\omega t) \, u_2\, \left( \epsilon_2 u_1 -\epsilon_1 u_2\right)\\
\delta v_1 &=&\frac{1}{4} \omega^2 \cos(\omega t) u_1 (2+u_2 v_1 -u_1 v_2)(\epsilon_1 u_2-\epsilon_2 u_1)\\
&&-\omega \sin(\omega t) \epsilon_1 ( 1+u_2 v_1-u_1 v_2) \\
\delta v_2 &=&\frac{1}{4} \omega^2 \cos(\omega t) u_2 (2+u_2 v_1 -u_1 v_2)(\epsilon_1 u_2-\epsilon_2 u_1)\\
&&-\omega \sin(\omega t) \epsilon_2 ( 1+u_2 v_1-u_1 v_2)
\end{array} \right\} . \label{Gtr2}
\eea
One may check that \p{ih2} as well as \p{eom2} are invariant under these transformations.

Can we invent an invariant action which yields the equations of motion~\p{eom2}?
The simplest candidate which produces \p{eom2} and is invariant under $M_0$ rotations 
(see~\p{m2}),
\be\label{testaction2}
S_{\textrm{test}} = \int dt \left( \dot{u}_1 \dot{u}_2 - \omega^2 \, u_1 u_2 \right)
\ee
is not invariant with respect to the other transformations in \p{conf2}, \p{m2} or \p{Gtr2}. 
For instance, under an $M_{-1}$ transformation (see again~\p{m2}) it changes by
\be\label{var}
\delta S_{\textrm{test}}= -  \alpha\, \int dt \left( \dot{u}_2 \dot{u}_2 - \omega^2\, u_2\,u_2\right) \neq 0.
\ee
In fact, with the given set of four coordinates $u_A$ and $v_A$ provided by the coset $\cW$ via \p{g2}) it is impossible to construct
an SO(2,3) invariant action. However, the variation of $S_{\textrm{test}}$ suggests that we introduce additional coordinates 
to compensate for the variation~\p{var}. These new variables must experience constant shifts under the $M_{-1}$ and $M_{1}$ transformations 
and carry the appropriate $M_0$ charge.
Therefore, the idea is to further extend our coset space from five to seven dimensions,
\be
\cW = \frac{\textrm{SO}(2,3)}{\textrm{SU}(1,1)\times\mathfrak{B}_{\textrm{SU}(1,1)}} \qquad \rightarrow\qquad 
\cW_{\textrm{imp}} = \frac{\textrm{SO}(2,3)}{\textrm{U}(1)\times\mathfrak{B}_{\textrm{SU}(1,1)}} ,
\ee
where the U(1) factor is generated by~$M_0$. 
The new Goldstone fields $\Lambda_{-1}$ and $\Lambda_{+1}$ associated with the generators $M_{-1}$ and $M_{+1}$, respectively, 
come with determined transformation properties.
Moreover, the Cartan form for the U(1) generator shifts by a time derivatives under any SO(2,3) transformation
\p{conf2}, \p{m2} or \p{Gtr2} and, therefore, may be considered for an invariant action.

To realize above mentioned procedure we have to perform the following steps.
\begin{itemize}
\item First, we must introduce the new ccordinates $\Lambda_{\pm1}$ by extending our coset element $g$ \p{g2} to 
\be\label{g2imp}
g_{\textrm{imp}} = g\; e^{ \Lambda_{-1} M_{-1} + \Lambda_{+1} M_+1}, \qquad \Lambda_{\pm 1}^\star = \Lambda_{\pm 1}.
\ee
\item Second, one has to recalculate the Cartan forms. Let us denote their `improved' version by $\Omega_{M_{a}}$ and $\Omega_{r,\alpha}$.
Then, the simplest invariant action is
\be\label{action22}
S= - \int \Omega_{M_0} .
\ee
\item Third, one has to derive the `improved' equations of motion for $u_1$ and $u_2$ from \p{action22}
and assert that they are unchanged, i.e.~still coincide with~\p{eom2}.
\end{itemize}
The improved Cartan forms are defined through the coset element $g_{\textrm{imp}}$ \p{g2imp} via
\be\label{cf2mod}
g_{\textrm{imp}}^{-1}\,d\,g_{\textrm{imp}} = \sum_{n} \Omega_{{L}_n}\, L_n+\sum_{a} \Omega_{{M}_a}\, M_a+\sum_{r,A} \Omega_{r,A} G_{r,A}
\ee
and read\footnote{The forms $\Omega_{L_n}=\omega_{L_n}$ are unchanged. We do not need to know their explicit form.}
\bea\label{mforms3}
&& \Omega_{\pm \frac{1}{2},-\frac{1}{2}} = \frac{1}{\sqrt{1+ \lambda_{-1} \lambda_{+1}}}\left(\omega_{\pm \frac{1}{2},-\frac{1}{2}} +
\lambda_{-1} \, \omega_{\pm \frac{1}{2},+\frac{1}{2}}\right), \nn \\
&& \Omega_{\pm \frac{1}{2},+\frac{1}{2}} = \frac{1}{\sqrt{1+ \lambda_{-1} \lambda_{+1}}}\left(\omega_{\pm \frac{1}{2},+\frac{1}{2}} -
\lambda_{+1} \, \omega_{\pm \frac{1}{2},-\frac{1}{2}}\right),  \nn \\
&& \Omega_{M_{-1}} = \frac{1}{1+\lambda_{-1} \, \lambda_{+1}}\left( d \lambda_{-1}  +\omega_{M_{-1}}+ \lambda_{-1} \omega_{M_0}+
\lambda_{-1}^2 \omega_{M_{+1}}\right), \nn \\
&& \Omega_{M_{+1}} = \frac{1}{1+\lambda_{-1} \, \lambda_{+1}}\left( d \lambda_{+1}  +\omega_{M_{{+1}}}- \lambda_{+1} \omega_{M_0}+
\lambda_{+1}^2 \omega_{M_{-1}}\right), \nn \\
&& \Omega_{M_0} =\frac{1}{1+\lambda_{-1}\lambda_{+1}}\left( 2 \lambda_{-1} \omega_{M_{+1}} - 2 \lambda_{+1} \omega_{M_{-1}} +
({1}-\lambda_{-1} \lambda_{+1})\,\omega_{M_0}+ \lambda_{-1} d\lambda_{+1} - \lambda_{+1} d \lambda_{-1}\right).
\eea
Here, stereographically projected coordinates were employed for simplicity,
\be\label{lambda}
\lambda_{-1} = \frac{ \tan\left( \sqrt{\Lambda_{-1} \Lambda_{+1}}\right)}{\sqrt{\Lambda_{-1} \Lambda_{+1}}} \Lambda_{-1} 
\qquad\textrm{and}\qquad
\lambda_{+1} = \frac{ \tan\left( \sqrt{\Lambda_{-1} \Lambda_{+1}}\right)}{\sqrt{\Lambda_{-1} \Lambda_{+1}}} \Lambda_{+1}.
\ee
The improved invariant constraints
\be\label{ih22imp}
\Omega_{-\frac{1}{2},\pm \frac{1}{2}} = 0 \qquad \textrm{and} \qquad \Omega_{+\frac{1}{2},\pm \frac{1}{2}} =0
\ee
imply the old constraints \p{ih2} and \p{eom2} and, therefore, indeed produce the previous equations of motion \p{eom2}.
For the new variables $\lambda_{\pm 1}$ one can get covariant equations of motion by imposing the extra constraints
\be\label{lamdih2}
\Omega_{M_{-1}} = \Omega_{M_{+1}} =0,
\ee
which imply
\be\label{lambdaeq2}
\dot{\lambda}_{-1} = \frac{ \left( \dot{u}_1 + \lambda_{-1}\, \dot{u}_2\right)^2}{2\left( 1{+}\frac{1}{2}\left(u_1 \dot{u}_2{-}\dot{u}_1 u_2  \right)\right)} -
\frac{\omega^2}{2} \left( u_1 + \lambda_{-1}\, u_2\right)^2,\quad
\dot{\lambda}_{+1} = \frac{ \left( \dot{u}_2 - \lambda_{+1}\, \dot{u}_1\right)^2}{2\left( 1{+}\frac{1}{2}\left(u_1 \dot{u}_2{-}\dot{u}_1 u_2  \right)\right)} -
\frac{\omega^2}{2} \left( u_2 - \lambda_{+1}\, u_1\right)^2.
\ee
Like the oscillator equations~\p{eom2}, the above are invariant under the transformations \p{conf2}, \p{m2}, \p{Gtr2}
together with the corresponding transformations of $\lambda_{-1}$ and $\lambda_{+1}$ The latter take the generic form
\be\label{lambdatr gen}
\delta\lambda_{-1} = \mu_{-1} + \mu_0 \lambda_{-1} + \mu_{+1} \lambda_{-1}^2 \qquad\textrm{and}\qquad
\delta\lambda_{+1} = \mu_{+1} - \mu_0 \lambda_{+1} + \mu_{-1} \lambda_{+1}^2,
\ee
with the parameters $\mu$ given by
\bea\label{lambdatr2}
g_0=e^{a\, L_{-1}+b\, L_0 +c\,L_{+1}}: &&
\mu_{-1} = \sfrac{1}{4} \ddot{f} u_1^2, \quad
\mu_0 = \sfrac{1}{2} \ddot{f} u_1\, u_2, \quad 
\mu_{+1} = \sfrac{1}{4} \ddot{f} u_2^2,
 \nn \\
g_0=e^{ \alpha \, M_{-1} + \beta\, M_0 + \gamma\, M_{+1}}: &&
\mu_{-1}= \alpha,\quad \mu_0 =\beta,\quad \mu_{+1}=\gamma,
 \nn \\[4pt]
g_0=e^{\epsilon_1\, G_{-\frac{1}{2},-\frac{1}{2}} + \epsilon_2 \,G_{-\frac{1}{2},+\frac{1}{2}}}: && 
\!\!\!\!\!\!\!\!\left\{ 
\begin{array}{l}
\mu_{-1}=\frac{1}{4} \omega^2\, \cos(\omega t)\, u_1^2 \left(\epsilon_1 u_2-\epsilon_2 u_1\right)-\omega \sin(\omega t)\,\epsilon_1\, u_1\\[2pt]
\mu_0=\frac{1}{2}  \omega^2\, \cos(\omega t)\, u_1\, u_2  \left(\epsilon_1 u_2-\epsilon_2 u_1\right)-
\omega \sin(\omega t)\left(\epsilon_1 u_2+\epsilon_2 u_1\right)\\[2pt]
\mu_{+1}=\frac{1}{4} \omega^2\, \cos(\omega t)\, u_2^2 \left(\epsilon_1 u_2-\epsilon_2 u_1\right)-\omega \sin(\omega t)\,\epsilon_2\, u_2
\end{array} \right\} . 
\eea

Finally, the invariant action \p{action22} acquires the form
\be\label{action22fin}
S =\int dt \left[ \frac{ \left( 1 -\lambda_{-1}\,\lambda_{+1}\right) \dot{u}_1 \dot{u}_2 +\lambda_{-1} \dot{u}_2^2 -\lambda_{+1} \dot{u}_1^2}
{\left( 1+\lambda_{-1}\,\lambda_{+1}\right)
\left( 1+\frac{1}{2}\left(u_1 \dot{u}_2- \dot{u}_1 u_2  \right)\right)}+
\frac{ \dot{\lambda}_{-1} \lambda_{+1} - \lambda_{-1} \dot{\lambda}_{+1}}{  1+\lambda_{-1}\,\lambda_{+1}}
-\omega^2 \frac{\left( u_1 +\lambda_{-1} u_2\right)\left( u_2- \lambda_{+1} u_1\right)}{1+\lambda_{-1}\,\lambda_{+1}}\right].
\ee
It is  matter of quite lengthly calculations to check the invariance of this action with respect to the transformations \p{conf2}, \p{m2}, \p{Gtr2} and
\p{lambdatr2}. A somewhat less tedious task is to check that the equations of motion following from the action \p{action22fin} coincide with the
equations \p{eom2} and \p{lambdaeq2}.

The action \p{action22fin} describes an interaction of the coordinates $u_1$ and $u_2$ with isospinor variables $\lambda_{-1}$ and $\lambda_{+1}$. 
Such kind of variables was firstly introduced within the supersymmetric Calogero model in~\cite{harm}. 
Somewhat later, these isospin variables (a.k.a.\ spin variables) were re-introduced through an SU(2)-reduction procedure~\cite{KL1,BKS1}. 
However, the action~\p{action22fin} has the following peculiarities, which distinguish it from a bosonic sector of some supersymmetric mechanics:
\begin{itemize}
\item We are dealing with the non-compact version of isospin variables, as they parametrize the coset $\textrm{SU}(1,1)/\textrm{U}(1)$.
Moreover, this SU(1,1) is not an external automorphism group but belongs to the symmetry of our system.
\item Despite the explicit interaction between isospin and ordinary variables in the action \p{action22fin}, 
the isospin variables decouple from $u_1$ and $u_2$ in the oscillator equations of motion~\p{eom2}.
They serve only to provide the SO(2,3) invariance of the action.
\end{itemize}
For an expected relationship of the action \p{action22fin} with those one constructed in \cite{BCN, BKNY}, one has to turn to the Hamiltonian formalism. 
This will be done elsewhere.

\setcounter{equation}{0}
\section{$G_{2(2)}$ invariant oscillator}
The 14-dimensional $g_{2(2)}$ algebra possesses a 5-grading with $d{=}4$ and again $\cH=su(1,1)$.
This is made manifest by its commutation relations,
\bea\label{G2}
&& \left[ L_n, L_m \right] = (n{-}m)\, L_{n+m}, \quad 
\left[ M_a, M_b \right] = (a{-}b)\, M_{a+b},\quad 
m,n =-1,0,1,\quad a,b =-1,0,1, \nn \\
&& \left[ L_n, G_{r,A}\right] = \left( \sfrac{n}{2}{-}r \right)  G_{n+r,A}, \quad
\left[ M_a, G_{r,A}\right] = \left( \sfrac{3 a}{2}{-}A \right)  G_{r,a+A},\quad 
r,s =-\sfrac12,\sfrac12, \quad A, B =-\sfrac32,-\sfrac12,\sfrac12,\sfrac32,\quad  \nn \\
&& \left[ G_{r,A}, G_{s,B} \right] =
3 A\left(4 A^2{-}5\right)  \delta_{A+B,0} \, L_{r+s} + \,r\left( 6 A^2{-}8 A\,B{+}6 B^2{-}9\right)
 \delta_{r+s,0} \, M_{A+B}.
\eea
As in the previous example \p{cr1}, these generators are chosen to be anti-hermitian,
\be\label{cr2}
\left( L_n\right)^\dagger = - L_n,\qquad 
\left( M_a\right)^\dagger = - M_{a},\qquad 
\left( G_{r,A}\right)^\dagger = - G_{r,A}.
\ee
Thus we have as basis elements
\be
G_{-\frac{1}{2}, A} \in \mathfrak{f}_{-\frac{1}{2}} , \qquad
G_{+\frac{1}{2}, A} \in \mathfrak{f}_{+\frac{1}{2}} \qquad\textrm{and}\qquad
M_a \in \cH = su(1,1).
\ee 
We start from the eight-dimensional quaternionic symmetric space 
$W=G_{2(2)}/\textrm{SO}(2,2)$ and enlarge it to the nine-dimensional coset
\be
\cW = \frac{G_{2(2)}}{\textrm{SU}(1,1)\times\mathfrak{B}_{\textrm{SU}(1,1)}}
\ee
with the stability subgroup generated by $(L_0,L_1,M_a)$ as before.
It may be parameterized as
\be\label{g3}
g= e^{t (L_{-1}+\omega^2 L_1)} 
e^{ u_1 G_{-\frac{1}{2},-\frac{3}{2}}+u_2  G_{-\frac{1}{2},-\frac{1}{2}}+u_3 G_{-\frac{1}{2},+\frac{1}{2}}+u_4  G_{-\frac{1}{2},+\frac{3}{2}}}
e^{ v_1 G_{+\frac{1}{2},-\frac{3}{2}}+v_2  G_{+\frac{1}{2},-\frac{1}{2}}+v_3 G_{+\frac{1}{2},+\frac{1}{2}}+v_4  G_{+\frac{1}{2},+\frac{3}{2}}},
\quad g^\dagger = g^{-1}.
\ee
The corresponding Cartan forms  are rather complicated. To write them in a concise form we re-label the generators $G$ and variables $u$ and $v$ 
in the spin-$\frac32$ $\cH$-representation with a symmetrized triple of spinor indices $\alpha,\beta,\gamma=1,2$:
\bea\label{newnotation}
&& G_{\pm \frac{1}{2},-\frac{3}{2}}=3 G_{\pm \frac{1}{2},111},\quad 
G_{\pm \frac{1}{2},-\frac{1}{2}}=3 G_{\pm \frac{1}{2},112},\quad
G_{\pm \frac{1}{2},+\frac{1}{2}}=3 G_{\pm \frac{1}{2},122},\quad
G_{\pm \frac{1}{2},+\frac{3}{2}}=3 G_{\pm \frac{1}{2},222}, \nn\\
&& u_1= \sfrac{1}{3} U^{111},\quad u_2 = U^{112},\quad u_3 = U^{122},\quad u_4 = \sfrac{1}{3} U^{222}, \nn \\
&& v_1= \sfrac{1}{3} V^{111},\quad\, v_2= V^{112},\quad v_3 = V^{122},\quad\, v_4 = \sfrac{1}{3} V^{222},
\eea
such that (with spinor index triples completely symmetric)
\bea
&& u_1 G_{-\frac{1}{2},-\frac{3}{2}}+ u_2 G_{-\frac{1}{2},-\frac{1}{2}}+ u_3 G_{-\frac{1}{2},+\frac{1}{2}}+ u_4 G_{-\frac{1}{2},+\frac{3}{2}} 
= \sum_{\alpha\beta\gamma} U^{\alpha\beta\gamma} G_{-\frac{1}{2},\alpha\beta\gamma} ,\nn \\
&& v_1\,G_{+\frac{1}{2},-\frac{3}{2}}+ v_2\,G_{+\frac{1}{2},-\frac{1}{2}}+ v_3\,G_{+\frac{1}{2},+\frac{1}{2}}+ v_4\,G_{+\frac{1}{2},+\frac{3}{2}}
= \sum_{\alpha\beta\gamma} V^{\alpha\beta\gamma} G_{+\frac{1}{2},\alpha\beta\gamma}.
\eea
Clearly, $G_{\pm \frac{1}{2},\alpha\beta\gamma}, U^{\alpha\beta\gamma}$, and  $V^{\alpha\beta\gamma}$ are real tensors 
totally symmetric in $\alpha,\beta,\gamma$.
Some of their multiple products will be abbreviated as follows,\footnote{ 
$su(1,1)$ indices are raised and lowered via
$\cA_{\alpha}=\epsilon_{\alpha\beta} \cA^\beta, \cA^{\alpha}=\epsilon^{\alpha\beta} \cA_\beta$ with 
$\epsilon_{\alpha\beta}\epsilon^{\beta\gamma} =\delta_\alpha^\gamma$ and $\epsilon_{12}=\epsilon^{21}=1$.}
\bea\label{short}
&& (\cA \cB)^{\alpha\beta} =\sfrac{1}{2} \sum \left( \cA^{\alpha\gamma_1\gamma_2}\cB_{\gamma_1\gamma_2}{}^{\beta} +
\cA^{\beta\gamma_1\gamma_2}\cB_{\gamma_1\gamma_2}{}^{\alpha}\right), \nn \\
&& (\cA \cB) =\sum \cA^{\alpha\beta\gamma} \cB_{\alpha\beta\gamma},\nn\\ 
&& (\cA \cB \cC)^{\alpha\beta\gamma}=\sfrac{1}{3}\sum\left( \cA^{\alpha\rho_1\rho_2}\cB_{\rho_1\rho_2\rho_3}\cC^{\rho_3 \beta\gamma}+
\cA^{\beta\rho_1\rho_2}\cB_{\rho_1\rho_2\rho_3}\cC^{\rho_3 \gamma\alpha}+
\cA^{\gamma\rho_1\rho_2}\cB_{\rho_1\rho_2\rho_3}\cC^{\rho_3 \beta\alpha}\right),\nn\\
&& (\cA \cB \cC \cD) =\sum \cA^{\alpha\beta_1\beta_2}\cB_{\beta_1\beta_2\gamma_1}\cC^{\gamma_1\rho_1\rho_2} \cD_{\rho_1\rho_2\alpha}.
\eea

Defining the Cartan forms
\be\label{cf3}
g^{-1} d g = \sum_{n} \omega_{{L}_n}\, L_n+\sum_{a} \omega_{{M}_a}\, M_a+
\sum_{\alpha\beta\gamma} \omega_u^{\alpha\beta\gamma}G_{-\frac{1}{2},\alpha\beta\gamma}+
\sum_{\alpha\beta\gamma} \omega_v^{\alpha\beta\gamma}G_{+\frac{1}{2},\alpha\beta\gamma},
\ee
we arrive at
\bea
\omega_u^{\alpha\beta\gamma}&=& dU^{\alpha\beta\gamma}+ \omega^2 dt \left( U^3\right)^{\alpha\beta\gamma}- V^{\alpha\beta\gamma}
\left[ dt \left( 1-\sfrac{\omega^2}{2}\left(U^4\right)\right)+ \left( UdU\right) \right], \label{cf3u} \\
\omega_v^{\alpha\beta\gamma}&=& dV^{\alpha\beta\gamma}+ (V^3)^{\alpha\beta\gamma}\left[dt\left( 1-\sfrac{\omega^2}{2}\left(U^4\right)\right)+ 
(UdU)\right]- 2  (VdUV)^{\alpha\beta\gamma} -(VVdU)^{\alpha\beta\gamma} \nn \\
&& +\;\omega^2 dt \Big[ U^{\alpha\beta\gamma}+3  (UUV)^{\alpha\beta\gamma}+
2 (VU^3V)^{\alpha\beta\gamma}-(U^3VV)^{\alpha\beta\gamma}\Big]. \label{cf3v}
\eea
In what follows we also need the forms $\omega_{M_a}$
\be\label{cf3m}
\omega_{M_{-1}}=\sfrac{1}{2} \omega^{11}, \qquad \omega_{M_{+1}}=\sfrac{1}{2} \omega^{22}, \qquad \omega_{M_{0}}= \omega^{12},
\ee
where
\be\label{omega}
\omega^{\alpha\beta} = -4  (VdU)^{\alpha\beta}+2  (VV)^{\alpha\beta}\left[ dt\left(1-\sfrac{\omega^2}{2} (U^4)\right)+ (UdU)\right]+
2 \omega^2 dt \Big[ (UU)^{\alpha\beta}+(U^3V)^{\alpha\beta}\Big].
\ee
Now, imposing the conditions $\omega_u^{\alpha\beta\gamma}=0$ we can express the coordinates $V^{\alpha\beta\gamma}$ in terms of $U^{\alpha\beta\gamma}$,
\be\label{ih33}
\omega_u^{\alpha\beta\gamma}=0\qquad \Rightarrow\qquad 
V^{\alpha\beta\gamma}=\frac{\dot{U}{}^{\alpha\beta\gamma}+ \omega^2  \left( U^3\right)^{\alpha\beta\gamma}}
{  1-\frac{\omega^2}{2}\left(U^4\right)+( U \dot{U})}.
\ee
With these relations the forms $\omega_{M_a}$ given by \p{cf3m} and \p{omega} simplify to
\be\label{cf3mF}
\omega_{M_{-1}}=\sfrac{1}{2} {\tilde\omega}^{11}, \qquad \omega_{M_{+1}}=\sfrac{1}{2} {\tilde\omega}^{22}, \qquad \omega_{M_{0}}= {\tilde\omega}^{12},
\ee
with
\be\label{omegaF}
{\tilde\omega}^{\alpha\beta} = -2 dt \frac{ (\dot{U}\dot{U})^{\alpha\beta}-\omega^2 \left[ \bigl(1+ (U\dot{U})\bigr)\,(UU)^{\alpha\beta}  -(U^3 \dot{U})^{\alpha\beta}\right]}
{  1-\frac{\omega^2}{2}\left(U^4\right)+( U \dot{U})}.
\ee
Finally, using the conditions $\omega_v^{\alpha\beta\gamma}=0$ in \p{cf3v} and the relations \p{ih33} we come to the covariant equations of motion
(with $V=V(U)$ according to~\p{ih33}):
\bea
&& \dot{V}{}^{\alpha\beta\gamma}+ (V^3)^{\alpha\beta\gamma}\left[\left( 1-\sfrac{\omega^2}{2}\left(U^4\right)\right)+ (U\dot{U})\right]-
 2  (V\dot{U}V)^{\alpha\beta\gamma} -(VV\dot{U})^{\alpha\beta\gamma} \nn \\
&& +\; \omega^2  \Big[ U^{\alpha\beta\gamma}+3  (UUV)^{\alpha\beta\gamma}+
2 (VU^3V)^{\alpha\beta\gamma}-(U^3VV)^{\alpha\beta\gamma}\Big] =0. \label{EOMFull}
\eea
In the limit $\omega=0$ these equations simplify to
\be\label{eom33}
\ddot{U}^{\alpha\beta\gamma} = 2 \frac{ (\dot{U}\dot{U}\dot{U})^{\alpha\beta\gamma} - \dot{U}{}^{\alpha\beta\gamma} (\dot{U}\dot{U}\dot{U}\cdot U)}{1+(U \dot{U})}
\qquad\textrm{with}\qquad
(\dot{U}\dot{U}\dot{U}\cdot U)\equiv \sum (\dot{U}\dot{U}\dot{U})^{\alpha_1\alpha_2\alpha_3}U_{\alpha_1\alpha_2\alpha_3},
\ee
and in the contraction limit $\gamma\to0$ after the rescaling $G{}^A_{\pm \frac{1}{2}} = \gamma^{-1}{\widetilde G}{}^A_{\pm \frac{1}{2}}$ (see~\p{confGal})
they linearize to
\be
\ddot{U}^{\alpha\beta\gamma} + \omega^2\,U^{\alpha\beta\gamma} = 0.
\ee

In full analogy with the SO(2,3) invariant oscillator considered in the previous section, 
in order to construct the invariant action one has to extend the coset to an eleven-dimensional one, 
\be
\cW = \frac{G_{2(2)}}{\textrm{SU}(1,1)\times\mathfrak{B}_{\textrm{SU}(1,1)}} \qquad \rightarrow\qquad 
\cW_{\textrm{imp}} = \frac{G_{2(2)}}{\textrm{U}(1)\times\mathfrak{B}_{\textrm{SU}(1,1)}} ,
\ee
with elements
\be\label{g3imp}
g_{\textrm{imp}} = g\, e^{\Lambda_{-1} M_{-1} + \Lambda_{+1} M_{+1}}
\ee
`improving' $g$ of \p{g3}. Defining the improved Cartan forms
\be\label{cf3imp}
g_{\textrm{imp}}^{-1} d g_{\textrm{imp}} = 
\sum_{n=-1}^{+1} \Omega_{{L}_n}\, L_n+\sum_{a=-1}^{+1} \Omega_{{M}_a}\, M_a+
\sum_{\alpha\beta\gamma}\Omega_u^{\alpha\beta\gamma}G_{-\frac{1}{2},\alpha\beta\gamma}+
\sum_{\alpha\beta\gamma}\Omega_v^{\alpha\beta\gamma}G_{+\frac{1}{2},\alpha\beta\gamma}
\ee
with $\Omega_{{L}_n}=\omega_{{L}_n}$,
one can see that $\Omega_u^{\alpha\beta\gamma}$ and $\Omega_v^{\alpha\beta\gamma}$ 
are linear combinations of the forms
$\omega_u^{\alpha\beta\gamma}$ and $\omega_v^{\alpha\beta\gamma}$, because
\bea
&&\Omega_u^{\alpha\beta\gamma}G_{-\frac{1}{2},\alpha\beta\gamma}=
e^{-\Lambda_{-1} M_{-1} - \Lambda_{+1} M_{+1}}\;
\omega_u^{\alpha\beta\gamma}G_{-\frac{1}{2},\alpha\beta\gamma}\;
e^{\Lambda_{-1} M_{-1} + \Lambda_{+1} M_{+1}}, \nn \\
&&\Omega_v^{\alpha\beta\gamma}G_{+\frac{1}{2},\alpha\beta\gamma}=
e^{-\Lambda_{-1} M_{-1} - \Lambda_{+1} M_{+1}}\;
\omega_v^{\alpha\beta\gamma}G_{+\frac{1}{2},\alpha\beta\gamma}\;
e^{\Lambda_{-1} M_{-1} + \Lambda_{+1} M_{+1}}.
\eea
Therefore, the analogous constraints on 
$\Omega_u^{\alpha\beta\gamma}$ and $\Omega_v^{\alpha\beta\gamma}$  
still imply the equations \p{ih33} and \p{EOMFull},
\be
\Omega_u^{\alpha\beta\gamma}=\Omega_v^{\alpha\beta\gamma}=0\qquad \Rightarrow \qquad
\omega_u^{\alpha\beta\gamma}=\omega_v^{\alpha\beta\gamma}=0.
\ee
The equations of motion for the additional variables $\lambda_{\pm1}$ related to $\Lambda_{\pm1}$ 
as in \p{lambda} follow from the invariant constraints
\bea\label{lambdaeom33}
&& \Omega_{M_{-1}} = \frac{1}{1+\lambda_{-1}\lambda_{+1}}
\left( d \lambda_{-1}  +\omega_{M_{-1}}+ \lambda_{-1} \omega_{M_0}
+\lambda_{-1}^2 \omega_{M_{+1}}\right) =0,
\nn \\
&& \Omega_{M_{+1}} = \frac{1}{1+\lambda_{-1}\lambda_{+1}}
\left( d \lambda_{+1}  +\omega_{M_{+1}}- \lambda_{+1} \omega_{M_0}
+\lambda_{+1}^2 \omega_{M_{-1}}\right) =0,
\eea
where the forms $\omega_{M_{-1}}, \omega_{M_{0}}$ and $\omega_{M_{+1}}$ 
were defined in \p{cf3mF} and \p{omegaF}.

Finally, the invariant action can be constructed from $\Omega_{M_0}$,
\be\label{action33}
S =-\int\, \Omega_{M_0} =-\int \frac{1}{1+\lambda_{-1}\lambda_{+1}}
\left[ \lambda_{-1} {\tilde\omega}^{22} - \lambda_{+1} {\tilde\omega}^{11} +
(1{-}\lambda_{-1} \lambda_{+1})\,{\tilde\omega}^{12}+
\lambda_{-1} d\lambda_{+1} - \lambda_{+1} d \lambda_{-1}\right],
\ee
where the ${\tilde\omega}^{\alpha\beta}$ were given in \p{omegaF}.

A good way to verify that the equations of motion extremize the action \p{action33} 
employs its first-order form
\be\label{action33F}
S =-\int\, \Omega_{M_0} =-\int \frac{1}{1+\lambda_{-1}\lambda_{+1}}
\left[ \lambda_{-1} {\omega}^{22} - \lambda_{+1} {\omega}^{11} +
(1{}-\lambda_{-1} \lambda_{+1})\,{\omega}^{12}+
\lambda_{-1} d\lambda_{+1} - \lambda_{+1} d \lambda_{-1}\right],
\ee
where the forms ${\omega}^{\alpha\beta}$ are given by the expressions \p{omega}. 
Then, varying this action over $V^{\alpha\beta\gamma}$ will yield~\p{ih33}, 
while the variations over $U^{\alpha\beta\gamma}$, $\lambda_{-1}$ and $\lambda_{+1}$ 
will reproduce $\omega_v^{\alpha\beta\gamma}=0$ and \p{lambdaeom33}, respectively.

The transformation properties of the time $t$ and the coordinates $U^{\alpha\beta\gamma}$
and $\lambda_{\pm 1}$ under $G_{2(2)}$ are found from computing the $G_{2(2)}$ action
on the improved coset elements~\p{g3imp} by left multiplication,
\be
g_0 \; g_{\textrm{imp}} = g'_{\textrm{imp}}\, h \qquad\textrm{with}\qquad g_0 \in G_{2(2)} 
\qquad\textrm{and}\qquad h \in \textrm{U}(1) \times \mathfrak{B}_{\textrm{SU}(1,1)}.
\ee
Due to the commutator relations \p{G2} it suffices to know the transformations generated by 
\be\label{g0g1}
g_0 = e^{\sum_{\alpha\beta\gamma}\epsilon^{\alpha\beta\gamma} G_{-\frac{1}{2}, \alpha\beta\gamma}} 
\qquad \textrm{and} \qquad
g_0 = e^{\sum_{\alpha\beta\gamma}\varepsilon^{\alpha\beta\gamma} G_{+\frac{1}{2}, \alpha\beta\gamma}}.
\ee
The corresponding transformations can be written in the following concise way,
\bea\label{FullG2tr}
&& \delta t = \frac{ (U^3 \vartheta) +(U \,\theta)}{1 +\frac{\omega^2}{2} (U^4)} , \qquad
\delta U^{\alpha\beta\gamma} = \theta^{\alpha\beta\gamma} +2 (U\vartheta U)^{\alpha\beta\gamma}+
(\vartheta U U)^{\alpha\beta\gamma} - \omega^2 (U^3)^{\alpha\beta\gamma} \delta t, \nn \\
&&\delta\lambda_{-1} = \Psi^{11} +2 \Psi^{12} \lambda_{-1}+ \Psi^{22} \lambda_{-1}^2, \qquad
\delta\lambda_{+1} =\Psi^{22} -2 \Psi^{12} \lambda_{+1} + \Psi^{11} \lambda_{+1}^2
\eea
with
\be
\Psi^{\alpha\beta} = (U \vartheta)^{\alpha\beta}- \omega^2 (U U)^{\alpha\beta} \delta t ,
\ee
and the parameters $\theta^{\alpha\beta}$ and $\vartheta^{\alpha\beta}$ are related to those in \p{g0g1} as
\be
\theta^{\alpha\beta\gamma} = 
\begin{cases} \cos(m t)\, \epsilon^{\alpha\beta\gamma} \\[2pt] 
\sfrac1m\sin(m t)\, \epsilon^{\alpha\beta\gamma} \end{cases}
\qquad\textrm{and}\qquad
\vartheta^{\alpha\beta\gamma} =
\begin{cases} -m\, \sin(m t)\, \epsilon^{\alpha\beta\gamma} \\[2pt] 
\cos(m t)\, \varepsilon^{\alpha\beta\gamma} \end{cases}
\ee
in the first and second instance of~\p{g0g1}, respectively.
A quite lengthy and tedious calculation confirms that the action \p{action33} is indeed invariant under these transformations.

\setcounter{equation}{0}
\section{Conclusions}
We proposed a procedure which associates with any simple Lie algebra a system of the second-order nonlinear differential equations which are
invariant with respect to a non-compact real form of this symmetry. Two explicit examples considered in detail gave rise to a system of deformed
oscillators invariant under SO(2,3) respective $G_{2(2)}$ transformations. For these cases, we also constructed invariant actions. 
These actions include additional, semi-dynamical variables which do not affect the equations of motion for the physical variables.

The five-graded decomposition of the Lie algebra, a key feature in our construction, coercively includes a one-dimensional conformal algebra~$su(1,1)$. 
Therefore, all systems constructed in this fashion will possess conformal invariance. Due to our special choice of the stability subalgebra 
a dilaton is absent, and the conformal invariance is achieved without it.
In a contraction limit, when the Lie algebra reduces to a Schr\"odinger algebra, the equations reduce to a system of ordinary harmonic oscillators.

The following further developments come to mind.\\[-16pt]
\begin{itemize}
\addtolength{\itemsep}{-4pt}
\item Our choice of the coset parametrization (the ordering $\mathfrak{g}_{-1}\cdot \mathfrak{g}_{-\frac{1}{2}} \cdot \mathfrak{g}_{\frac{1}{2}}$)
is rather special. Clearly, this is far from unique, and a reordering will give the equations a different appearance.
\item The chosen coset parametrization is calculationally useful but provides an unusual form of the metric. 
It is desirable to bring the metric and connection to a more standard form through some reparametrization.
\item Some Lie algebras possess other forms of grading (for example, there is a 7-graded basis for $G_2$).
It will be interesting to learn how our equations change when the grading is altered.
\item Our construction procedure for invariant actions works properly only in the presence of an $su(1,1)$ factor in the stability subalgebra. 
It should be clarified how to construct invariant actions when this is not so.
\item A supersymmetric extension of the present approach may be of interest.
\item Finally, a Hamiltonian description may illuminate the structure of conserved currents and help to relate our systems to others in the literature.
\end{itemize}
\section*{Acknowledgements}
The work of S.K.\ was partially supported by
RFBR grant 15-52-05022 Arm-a and the Heisenberg-Landau program.
The work of O.L.\ and of A.S.\ was partially supported by DFG grant Le-838/12-2.
The work of A.S.\ was partially supported also by
RFBR grant 15-52-05022 Arm-a, RFBR grant 16-52-12012-NNIO-a and the Heisenberg-Landau program.
This article is based upon work from COST Action MP1405 QSPACE,
supported by COST (European Cooperation in Science and Technology).

\end{document}